\documentclass{iaus}
\usepackage{graphicx}

\title[The diffuse X-ray emission of Planetary Nebulae] 
{Modeling the diffuse X-ray emission of Planetary Nebulae
with different chemical composition}

\author[M. Steffen et al.]   
{M. Steffen$^1$, C. Sandin$^1$,  R. Jacob$^1$, D. Sch\"onberner$^1$
}

\affiliation{$^1$Leibniz-Institut f\"ur Astrophysik Potsdam (AIP), \\ 
An der Sternwarte 16, 14482 Potsdam, Germany
}

\pubyear{2011}
\volume{283}  
\pagerange{101--104}
\jname{Planetary Neblae, an Eye to the Future}
\editors{A. Manchado, L. Stanghellini \& D. Sch\"onberner, eds.}
\begin{document}

\maketitle

\begin{abstract}
Based on time-dependent radiation-hydrodynamics simulations of the evolution
of Planetary Nebulae (PNe), we have carried out a systematic parameter study
to address the non-trivial question of how the diffuse X-ray emission of PNe
with closed central cavities is expected to depend on the evolutionary state
of the nebula, the mass of the central star, and the metallicity of
stellar wind and circumstellar matter. We have also investigated how the model
predictions depend on the treatment of thermal conduction at the interface
between the central `hot bubble' and the `cool' inner nebula, and compare the
results with recent X-ray observations. Our study includes models whose
properties resemble the extreme case of PNe with Wolf-Rayet type central 
stars. Indeed, such models are found to produce the highest X-ray luminosities.

\end{abstract}

\firstsection 
\section{Introduction}
Over the past decade, roughly a dozen nearby Galactic Planetary Nebulae were
found to be a source of extended diffuse X-ray emission (e.g.\ 
\cite[Kastner et al.\ 2008]{kastner2008}; \cite[Kastner 2009]{kastner2009}). 
Images taken from space by the X-ray observatories \emph{Chandra} and 
\emph{XMM-Newton} reveal that the diffuse X-ray emission is confined to 
the closed central cavities of round and elliptical PNe, indicating the 
presence of hot gas with temperatures in excess of $10^6$~K (for recent 
examples see Guerrero et al., this volume). Basically, these findings confirm 
theoretical predictions of wind-heated `hot bubbles', even though the observed 
X-ray temperatures are much lower (at least a factor of $10$) than obtained 
from simple spherical models assuming that the wind kinetic energy is
thermalized in an adiabatic shock. However, more elaborate models taking into 
account the heat flux due to thermal conduction across the central `hot bubble',
show much lower X-ray temperatures, in good agreement with observational 
evidence (\cite[Steffen et al.\ 2008]{steffen2008}, henceforth Paper~V). 
About $75\%$ of the targeted PNe show no evidence of diffuse X-rays, among
them old, evolved nebulae and objects with open structures. Note that a large
fraction of PNe also harbor X-ray point sources at the position of their
central stars.  This type of X-ray emission, however, is of a different nature
than the extended X-ray emission from the shock-heated hot wind bubbles which
are the subject of the present study.

The number of PNe with detected diffuse X-ray emission is too small to allow
the derivation of reliable empirical trends, such as relations between stellar
wind parameters and X-ray properties of the nebula. Likewise, it is difficult
to make theoretical predictions based on simple back-of-the-envelope
calculations. In very general terms, the X-ray luminosity, $L_{\rm X}$, is 
given by the X-ray emissivity per unit mass and electron density, $\eta$, 
integrated over the emitting volume: 
$L_{\rm X}=\overline{\eta}(T_{\rm e},Z)\,M_{\rm X}\,\overline{n_{\rm e}}
\sim \overline{\eta}(T_{\rm e},Z)\,M_{\rm X}^2\,R_{\rm X}^{-3}$. The
emissivity depends on the electron temperature, $T_{\rm e}$, and the 
metallicity of the plasma, $Z$; the mean electron density is proportional to 
the X-ray emitting mass, $M_{\rm X}$, divided by the X-ray emitting volume, 
$\overline{n_{\rm e}}\sim M_{\rm X}\,R_{\rm X}^{-3}$. If we want to 
know how the X-ray luminosity of an object changes in the course of its
evolution, we have to know how the product $M_{\rm X}^2\,R_{\rm X}^{-3}$
changes with time. This depends mainly on the time evolution of the fast wind 
of the central star: for a constant wind expanding into an environment with a
radial density profile $\rho \propto r^{-2}$, it can be shown that 
$M_{\rm X} \propto t$, and $R_{\rm X} \propto t$ 
(\cite[Koo \& McKee 1992]{koo1992}), such that the X-ray luminosity 
\emph{decreases} as $L_{\rm X} \propto t^{-1}$.
On the other hand, the more realistic assumption that wind power 
and momentum flux increase with time, e.g.\, $\dot{M}\,V^2 \propto t^{\,2.62}$, 
$\dot{M}\,V \propto t^{\,1.02}$, leads to an \emph{increasing} X-ray
luminosity, $L_{\rm X} \propto t^{\,1.2}$ (\cite[Zhekov \& Perinotto
1996]{zhekov1996}).
Moreover, properties of the X-ray emission also depend on the thermal
structure of the hot plasma, which is ruled by the efficiency
of electron heat conduction. Clearly, detailed numerical models are 
necessary to make reliable theoretical predictions.

\section{Model calculations}
Using the Potsdam NEBEL code, we model the combined evolution of central star
(CS) and circumstellar envelope by radiation hydrodynamical simulation in 1D
spherical geometry, accounting for non-equilibrium ionization of the nine most
important chemical elements. A typical model has a radial extent from
$6\times 10^{14}$ to $3\times 10^{18}$~cm ($0.0002$ to $1.0$~pc). Treated in 
a physically consistent way, it contains the freely expanding CS wind, the
inner reverse shock, the `hot bubble' consisting of the shocked stellar wind, 
and the (double-shell) PN proper, separated from the `hot bubble' by the contact
discontinuity and from the surrounding AGB halo by the outer shock.  
Temporally, the simulations start at the tip of the AGB and are advanced 
well into the white dwarf regime, thus covering the formation and complete 
evolution of the PN.

Given the radial temperature and density structure of the
hydrodynamical model, we apply the \emph{Chianti} code 
(v6.0.1, \cite[Dere et al. 1997, 2009]{d97,d09}) to compute, for
selected instants along the evolutionary sequence, the emergent X-ray
spectrum, the X-ray luminosity, $L_{\rm X}$, the characteristic X-ray 
temperature, $T_{\rm X}$, and the surface brightness distribution of the X-ray
emitting hot central cavity. For details about the model calculations see
Paper~V and references therein.

\section{Simulated X-ray emission: a parameter study}
The work by \cite{steffen2008} was restricted to models with standard
Galactic-disk elemental abundances, as given in their Table\,1. The present
parameter study considers the influence of the adopted chemical composition 
on the resulting X-ray properties. 

\emph{Models with scaled metallicities.}
We have obtained model sequences of different metallicity, $Z$, by globally
scaling the number densities (relative to H) of the 'metals' 
C, N, O, Ne, S, Cl, and Ar by a common factor $f_Z$ with respect to the
standard Galactic-disk abundance mix, $Z=f_Z\,Z_{\rm GD}$. 
As explained in detail in 
\cite{schoenberner2010}, changes of the metal content affect both the 
cooling properties of the nebular matter (higher $Z \Rightarrow$ more
efficient line cooling), and the wind of the central star (higher $Z 
\Rightarrow$ higher wind power). Specifically, we assume here $\dot{M}
\propto Z^{0.69}$ and $V_{\rm wind} \propto Z^{0.13}$, so that $L_{\rm wind}
\equiv 1/2\,\dot{M}\,V_{\rm wind}^2 \propto Z^{0.95}$. At the same time, we 
assume that the evolution of the central star, $T_{\rm eff}(t)$, 
$L_\ast(t)$, is unaffected by changes of $Z$
(see \cite[Sch\"onberner at al.\ 2010]{schoenberner2010}).

\begin{figure}[t]
\begin{center}
 \mbox{\includegraphics[bb=50 40 550 360,width=10cm]{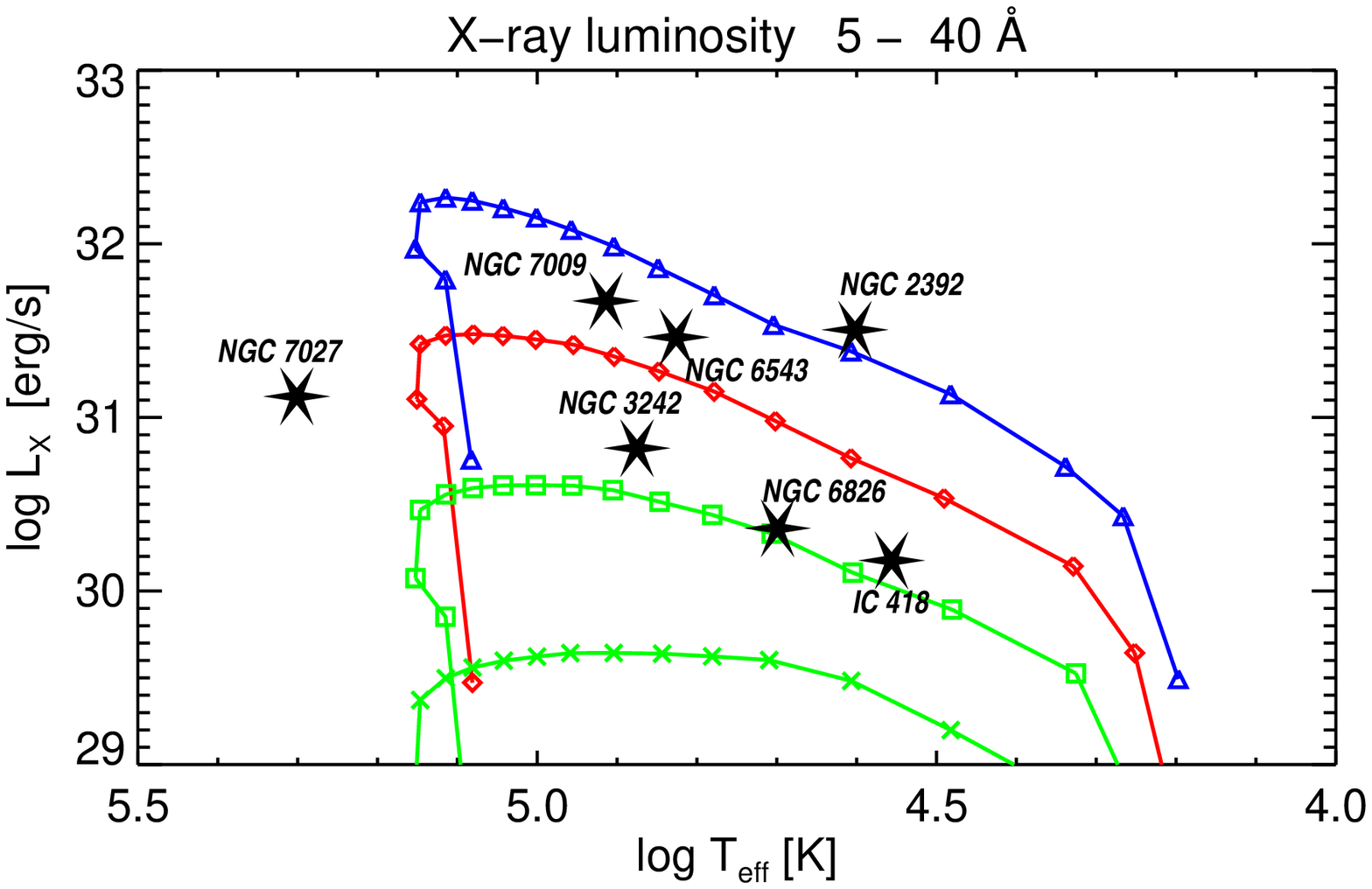} }
 \mbox{\includegraphics[bb=50 40 550 360,width=10cm]{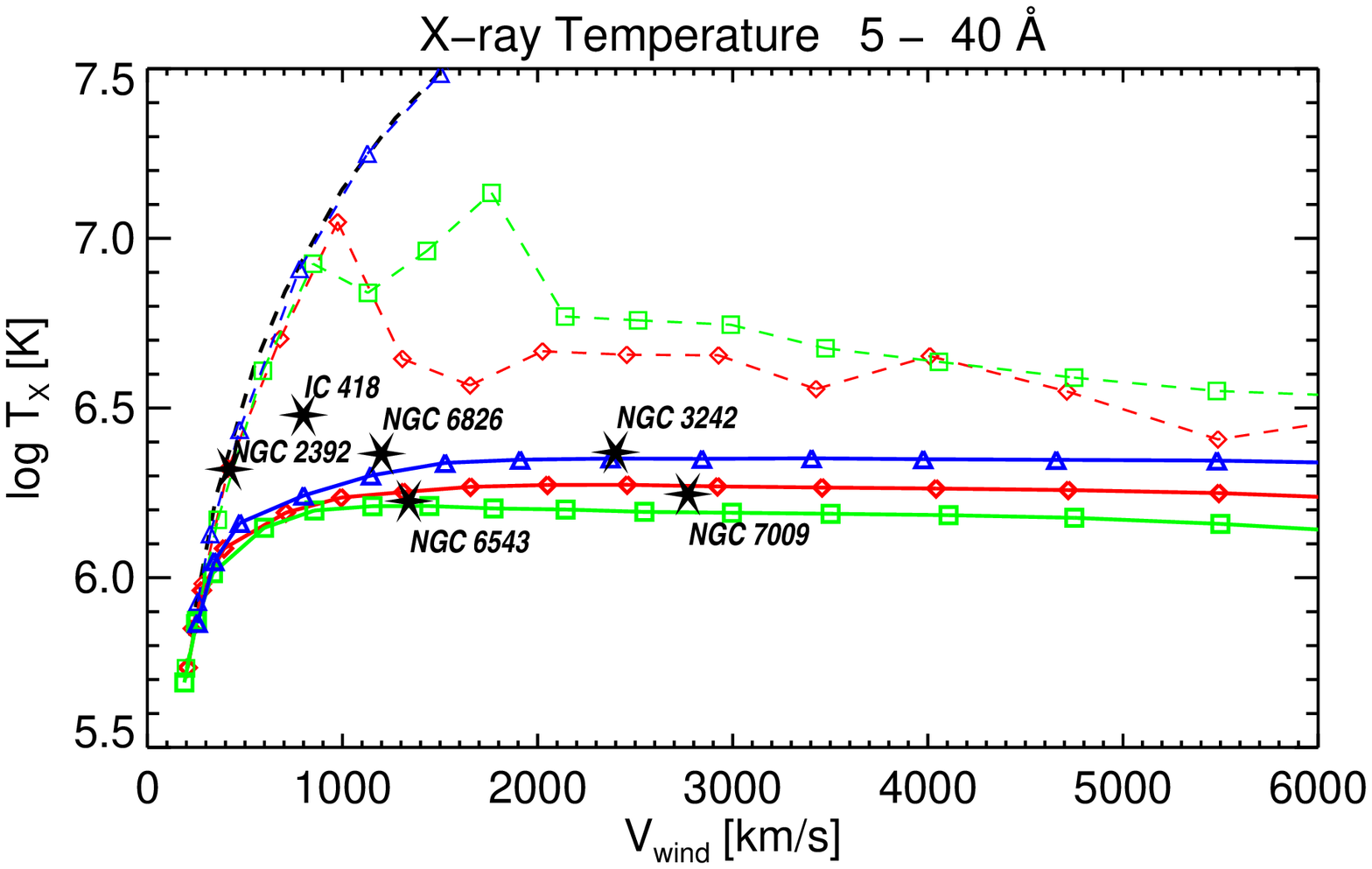} }
 \vspace*{0.5 cm}
 \caption{Evolution of the X-ray luminosity, $L_{\rm X}$, as a function of 
the CS effective temperature (left), and of the X-ray temperature, $T_{\rm
  X}$, as a function of the CS wind speed (right), for a central star with
$M_\ast=0.595$~M$_\odot$.  The solid curves connecting data points (symbols)
refer to model sequences computed with heat conduction HC2 (see Paper~V),
assuming metallicity scaling factors of $f_Z=3$ (triangles), $1$ (diamonds),
$1/3$ (squares), and $1/10$ (crosses). Corresponding sequences computed
without heat conduction are shown by the dashed curves (right panel only),
omitting the data points for $f_Z=3, V_{\rm wind}>2000$~km/s which are
numerically uncertain. Stars mark the position of observed PNe with normal
(H-rich) chemical composition (data M.~Guerrero 2011, priv. comm.). $L_{\rm
  X}$ and $T_{\rm X}$ refer to the X-ray emission between $5$ and
$40$\,\AA\ ($0.3$ to $2.5$~keV).}
\label{fig1}
\end{center}
\end{figure}

Our previous studies at standard Galactic-disk composition, based on models
including heat conduction, had shown that the X-ray luminosity increases 
steadily  with time ($T_{\rm eff}$) until the turnaround point (maximum 
$T_{\rm eff}$) is reached, after which $L_\ast$, $L_{\rm wind}$, and $L_{\rm X}$ drop
rapidly. Essentially, the time evolution of $L_{\rm X}$ reflects the evolution of
$L_{\rm wind}$. The so far unexplored \emph{metallicity dependence} of 
$L_{\rm X}$ and $T_{\rm X}$ is shown in Fig.\,\ref{fig1} for a central star with 
$M_\ast=0.595$~M$_\odot$. For given $T_{\rm eff}$, $L_{\rm X}$ increases
systematically with metallicity (left panel). Naively, one might have expected 
that  roughly $L_{\rm X} \propto Z^2$, since both $L_{\rm wind}$ and the number of 
metal ions scale with $Z$. The numerical results, however, show a weaker $Z$ 
sensitivity, $L_{\rm X} \propto Z^p, 1 < p < 2$, where $p$ depends on 
$T_{\rm eff}$.  The X-ray temperature increases only slightly with metallicity
(right panel), in quantitative agreement with the scaling suggested by
Eq.\,(14) in Paper~V, noting that $L_{\rm wind} \propto Z$. In the absence of
heat conduction, $T_{\rm X}$ is significantly higher (Fig.\,\ref{fig1}, right
panel) and -- except for the earliest stages of evolution -- $L_{\rm X}$ is
reduced by roughly 2 orders of magnitude (not shown).

We have also investigated sequences of different metallicity for 
$M_\ast=0.625$~M$_\odot$ and $0.696$~M$_\odot$. Qualitatively, the $Z$
dependence of the X-ray properties is very similar to the case 
$M_\ast=0.595$~M$_\odot$. Summarizing the combined
$Z$, $M_\ast$, $t$ dependence, the models predict the highest $L_{\rm X}$
for metal-rich nebulae with evolved (hot) massive central stars. 

\begin{figure}[t]
\begin{center}
 \mbox{\includegraphics*[bb=50 28 570 342,width=10cm]{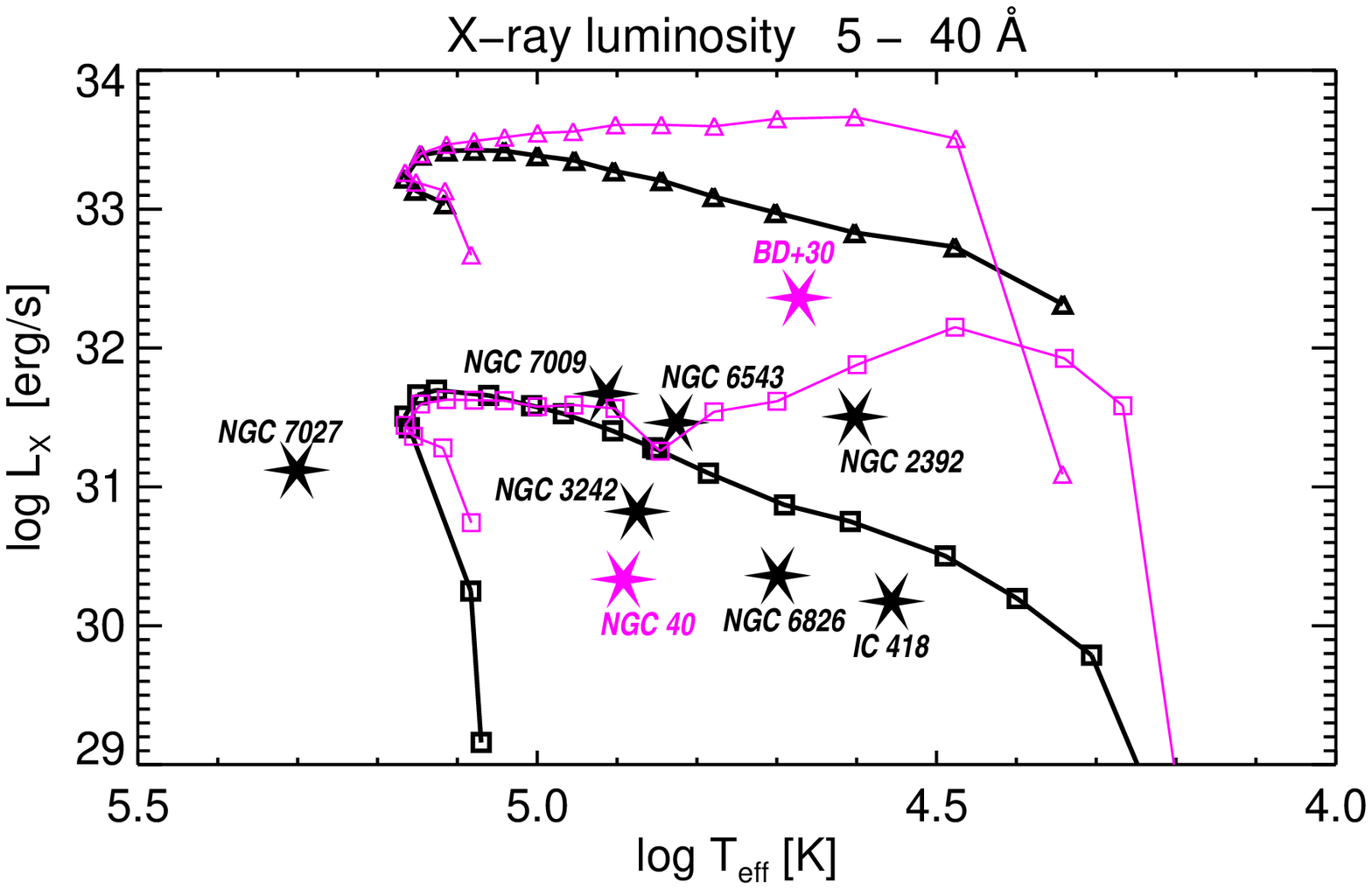} }
 \mbox{\includegraphics*[bb=20 28 570 365,width=10cm]{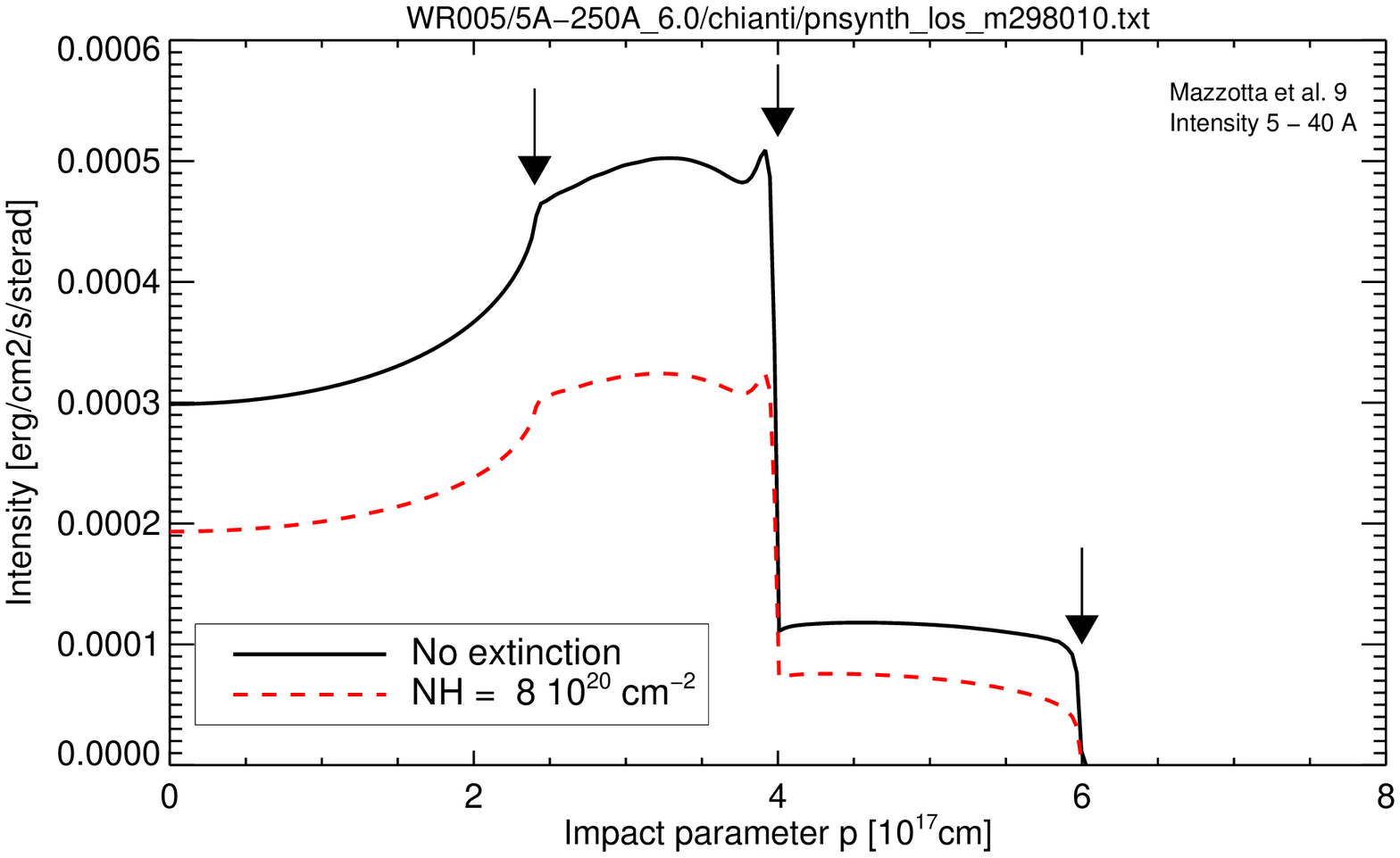} }
 \caption{{\bf Left:} Evolution of $L_{\rm X}$ over CS effective 
temperature, for $M_\ast=0.595$~M$_\odot$.  The two thick curves refer to
model sequences with standard chemical composition, the corresponding thin 
curves show the effect of changing the CS wind chemistry from standard
(H-rich) to  WR-type (H-deficient) at approximately identical $\dot{M}(t)$, 
$V_{\rm wind}(t)$, and unchanged H-rich composition of the nebula. 
The upper two tracks (triangles) have $\dot{M}(t)$ increased by a factor 
$100$ ($V_{\rm wind}(t)$ unchanged) with respect to the standard mass loss 
rate used with the lower two tracks (squares). All sequences
have been computed with heat conduction HC2 (see Paper~V). The observed 
positions of two PNe with WR-type central stars are also indicated (stars
marked BD$+30^\circ3639$ and NGC~$40$; data M.~Guerrero 2011, priv. comm.).  
{\bf Right:} Synthetic X-ray surface brightness profile computed for the
uppermost track in the left panel at $T_{\rm eff}\approx 90\,000$~K. From left
to right, arrows indicate the position of the inner shock, the chemical
discontinuity, and the conduction front. Both plots refer to the X-ray
emission between $5$ and $40$\,\AA\ ($0.3$ to $2.5$~keV).}
\label{fig2}
\end{center}
\end{figure}

\emph{Models with Wolf-Rayet-type central stars.}
Wolf-Rayet (WR) type central stars are characterized by a hydrogen-deficient
chemical composition of their atmospheres and winds, and by mass loss 
rates that exceed those of normal central stars by factors $10$ to $100$.
In a first attempt to model the X-ray emission of PNe with WR-type
central stars, we have modified our standard simulation 
($M_\ast$=$0.595$~M$_\odot$) in two respects. First, we changed the chemical 
composition of the CS wind from standard (H-rich) to WR-type (2\% H, 42\% He, 
49\% C, 5\% O by mass) at the tip of the AGB ($t$=$0$). This implies that the
`hot bubble' forms out of H-deficient matter, while the surrounding nebula
remains H-rich.  For consistency, we have updated the NEBEL code with a
generalized description of electron heat conduction that is valid for
arbitrary chemical composition (Sandin et al., this volume).
The impact of changing the chemistry of the CS wind is shown by the 
difference between the two lower curves in Fig.\,\ref{fig2} 
(left panel). The H-deficient `hot bubble' forms at
$t$$=$$1340$\,y, $T_{\rm eff}$=$15.4$\,kK, $V_{\rm wind}$=$260$\,km/s, slightly
later than the H-rich bubble ($1190$\,y, $13.1$\,kK, $165$\,km/s), but 
reaches much higher X-ray luminosities (up to a factor $100$) during the 
early evolution (low $T_{\rm eff}$). Obviously, the X-ray emissivity (per unit 
mass) is much higher for WR-type matter. Later, the difference in $L_{\rm X}$ 
tends to vanish, since the bulk of the hot gas is increasingly made up of 
evaporated H-rich matter (see below). As a second step, we have increased 
the CS mass loss rate globally by a factor $100$. This delays the formation
of the `hot bubble' both for normal ($1620$\,y, $19.3$\,kK, $325$\,km/s), and
WR-composition CS winds ($2160$\,y, $25.4$\,kK, $720$\,km/s). As seen in 
Fig.\,\ref{fig2} (left, upper two curves), such models easily exceed 
the X-ray luminosity measured for BD$+30^\circ3639$, a well studied PN with a
[WC]-type central star. Again, corresponding models calculated
without heat conduction typically have lower $L_{\rm X}$ by roughly a factor
$100$, except at the very early stages where heat conduction is negligible.
Finally, Fig.\,\ref{fig2} (right) shows the X-ray surface
brightness distribution of the test model with WR-type wind composition
and scaled-up mass loss rate at a late stage of evolution 
($T_{\rm eff}\approx 90\,000$~K). The `hot bubble' is divided into two
chemically distinct parts: the inner region ($r\le 4\times 10^{17}$~cm) 
consists of H-deficient matter, the outer shell consists of H-rich 
matter that evaporated due to heat conduction. In X-rays, the H-deficient
core of the central cavity shines much brighter than the surrounding
H-rich layer (cf.\ Jacob et al., this volume).

\section{Conclusions}
The observed X-ray luminosities of PNe with normal H-rich composition
can be explained by hydrodynamical models with heat conduction, 
either assuming a standard Galactic-disk metallicity, $Z_{\rm GD}$, and a
range of central star masses between $M_\ast=0.565$ and $0.7$~M$_\odot$, 
or by a fixed $M_\ast=0.625$~M$_\odot$ and a range of metallicity scaling 
factors between $f_Z=1/3$ and $f_Z=3$. The highest X-ray fluxes are expected 
to originate from PNe with WR-type central stars, due to their powerful and 
H-deficient winds. Models with suppressed thermal conduction (magnetic fields) 
may explain X-ray non-detections.

%


\begin{thebibliography}{}

\bibitem[Dere \etal\ (1997)]{d97}
{Dere, K.P., Landi, E., Mason, H.E., Fossi, B.C., Young, P.R.} 1997, 
\textit{A\&AS}, 125, 149

\bibitem[Dere \etal\ (2009)]{d09}
{Dere, K.P., Landi, E., Young, P.R., Del Zanna, G., Landini, M., 
Mason, H.E.} 2009, 
\textit{A\&A}, 498, 915

\bibitem[Kastner et al.(2008)]{kastner2008} 
{Kastner, J.H., Montez, R., Balick, B., De Marco, O.} 2008, 
\textit{ApJ}, 672, 957 

\bibitem[Kastner (2009)]{kastner2009}
{Kastner, J.} 2009, in: S. Wolk, A. Fruscione, D. Swartz (eds.), 
\textit{Chandra's First Decade of \mbox{Discovery}}, 
http://cxc.harvard.edu/symposium\_2009/proceedings/session\_01.html$\#$talk03

\bibitem[Koo \& McKee (1992)]{koo1992}
{Koo, B.-C., McKee, C.F.} 1992,
\textit{ApJ}, 338, 103

\bibitem[Sch\"onberner et al. (2010)]{schoenberner2010}
{Sch\"onberner, D., Jacob, R., Sandin, C., Steffen, M.} 2010,
\textit{A\&A}, 523, A86

\bibitem[Steffen et al. (2008)]{steffen2008}
{Steffen, M., Sch\"onberner, D., Warmuth, A.} 2008,
\textit{A\&A}, 489, 173 (Paper~V)

\bibitem[Zhekov \& Perinotto (1996)]{zhekov1996}
{Zhekov, S.A., Perinotto, M.} 1996,
\textit{A\&A}, 309, 648


\end{thebibliography}
\end{document}